\documentclass[aps,prl,twocolumn,showpacs,amsmath]{revtex4}
\usepackage{graphicx}
\begin{document}
\title{Large magnetic field-induced spectral weight enhancement of high-energy spin excitations in
${\bf La_{1.88}Sr_{0.12}CuO_{4}}$}
\author{L.~H.~Machtoub$^{1}$, B.~Keimer$^{1}$, and K.~Yamada$^{2}$}
\affiliation{$^{1}$Max Planck Institute for Solid State Research,
Heisenbergstr. 1, D-70569 Stuttgart, Germany\\
$^{2}$Institute for Material Research, Tohoku University, Katahira 2-1-1, Sendai, 980-8577,
Japan}
%\email{L.Machtoub@fkf.mpg.de}
\begin{abstract}
We report electronic Raman scattering experiments on a
superconducting ${\rm La_{1.88}Sr_{0.12}CuO_{4}}$ single crystal
in a magnetic field. At low temperatures, the spectral weight of
the high-energy two-magnon peak increases linearly with field and
is amplified by a factor of more than two at 14 T. The effect
disappears at elevated temperatures and is not present in undoped
${\rm La_{2}CuO_{4}}$. This observation is discussed in terms of
an electronically inhomogeneous state in which the field enhances
the volume fraction of a phase with local antiferromagnetic order
at the expense of the superconducting phase.
\end{abstract}


\maketitle

Shortly after the discovery of high temperature superconductivity,
antiferromagnetic Mott insulating and superconducting states were
shown to be directly adjacent in the phase diagram of the doped
copper oxides. This early observation still serves as primary
evidence of the prominent role of Coulomb correlations in the
mechanism of high temperature superconductivity. Recent neutron
scattering \cite{katano,lake_nature,khaykovitch}, nuclear magnetic
resonance \cite{mitrovic,kumagai}, and muon spin rotation
\cite{kiefl} experiments have indicated that the interplay between
these two states is more delicate than had long been assumed. In
underdoped copper oxides with hole concentrations per copper atom,
x, around $1/8$, an external magnetic field much lower than the
upper critical field of the superconducting state was shown to
induce or enhance static antiferromagnetic order. In a more highly
doped sample with $x \sim 0.16$, a field of 5T was observed to
strongly modify the spin excitations spectrum around 4 meV, an
energy that significantly exceeds the Zeeman energy
\cite{lake_science}. Theoretical work has attributed these
findings to an enhancement of either static (for $\rm x \sim 1/8$)
or dynamic (for $\rm x > 1/8$) spin correlations around the vortex
cores of the superconductor
\cite{arovas,ogata,ting,zhang,franz,berlinsky,bishop,hedegard}.
The 4 meV mode observed in the experiments of Ref.
\onlinecite{lake_science} has been interpreted as a soft mode
associated with a putative quantum critical point separating the
superconducting state from a state with magnetic long-range order
\cite{sachdev}. In this picture, its field-induced spectral weight
amplification results from an amplitude enhancement of the mode
inside the vortex core.

Here we use magnetic Raman scattering to probe the magnetic field
dependence of spin excitations at much larger energies, comparable
to the antiferromagnetic superexchange coupling $J \sim 100$ meV.
In insulating, antiferromagnetically ordered cuprates, a prominent
two-magnon excitation peak is observed in Raman spectra at an
energy of $\sim 2.3 J$ \cite{sugai}. The spectral weight of this
peak is dominated by local spin-flip excitations near the
Brillouin zone boundary of the antiferromagnet. Upon doping, the
two-magnon peak broadens and ultimately merges into the charge
excitation continuum of the conduction electrons \cite{sugai}.
Here we show that the spectral weight of the two-magnon peak in
the superconducting state of ${\rm La_{2-x}Sr_{x}CuO_{4}}$ with
$\rm x \sim 1/8$ is enhanced by a factor of more than two in a
magnetic field of 14 T. This profound spectral weight
renormalization of the highest-energy, local spin flip excitations
cannot be understood by the soft-mode behavior invoked to explain
the field dependence of the long-wavelength spin excitations
observed in low-energy spectroscopies. It thus requires a revision
of our understanding of the interplay between antiferromagnetic
order and superconductivity in underdoped copper oxides. Possible
implications will be discussed below.

A single crystal of ${\rm La_{1.88}Sr_{0.12}CuO_{4}}$ was grown by
the travelling solvent floating zone technique as described
previously \cite{katano}. Its superconducting transition
temperature, $\rm T_c = 34$ K, was determined by monitoring the
diamagnetic response in a magnetometer. An undoped ${\rm
La_{2}CuO_{4}}$ crystal was grown by the same method and annealed
under Ar flow at 950$^\circ$C in order to remove oxygen
interstitials. After the annealing procedure, its N\'eel
temperature (also determined by magnetometry) was 320 K. The Raman
scattering measurements were performed in quasi-back scattering
geometry using a triple monochromator (Dilor $xy$), a charge
coupled device (CCD) detector, and a laser wave length of 514.5
nm. The scattered light was collected along the crystallographic
$c$-axis. The magnetic field dependent data were obtained using a
superconducting magnet (Oxford Instruments) with a maximum field
of 14 T, operated in a temperature range from 4.2 to 300 K. The
samples were mounted in a continuous helium flow cryostat
installed in the magnet, and the field was applied perpendicular
to the ${\rm CuO_{2}}$ planes. The beam spot on the sample surface
was monitored by a specially designed electronic sensor and a
camera installed inside the spectrometer. By adjusting the sample
position to compensate for magnetostriction and thermal expansion
of the sample mount, it proved possible to keep the beam position
fixed during field and temperature changes, thereby minimizing
systematic errors due to variations in surface morphology. Further
experimental precautions included field-cooling the samples
through the superconducting transition, in order to avoid strains
due to flux trapping. The laser power was about 10W/${\rm
cm^{2}}$. By examining the intensity ratio of the Stokes and
anti-Stokes spectra, the temperature of the illuminated region of
the sample was estimated to be less than 5 K at low temperature.
Spectral corrections were made for the frequency dependence of the
collection optics and spectrometers, as well as the detector
sensitivity. The spectra were taken in $x'y'$ geometry, that is,
with the incident and final photon polarization states at an angle
of 45$^\circ$ from the Cu-O bonds. (The small orthorhombic
distortion of the copper oxide layers can be ignored.) This yields
Raman spectra of ${\rm B_{1g}}$ symmetry, where the two-magnon
peak has the largest amplitude.

Figures 1 and 2 show ${\rm B_{1g}}$ Raman spectra collected for
the ${\rm La_{1.88}Sr_{0.12}CuO_{4}}$ crystal at different
temperatures and magnetic fields. The data in zero field are in
excellent agreement with prior work on similar samples
\cite{sugai,naeini}. Upon cooling, they exhibit a gradual
evolution from a single broad peak around 3000 cm$^{-1}$ at room
temperature to a two-peak profile at base temperature. The two
peaks in the low temperature profile are also seen in the
electronic Raman spectra of doped nickelates \cite{blumberg2},
where phases with ``stripe" order of spin and charge are well
documented. They have thus been ascribed to two-magnon excitations
propagating along and perpendicular to the charged domain walls of
a ``striped" phase \cite{sugai}.

\begin{figure}[t]
%h=here, t=top, b=bottom, p=separate figure page
\begin{center}\leavevmode
\center{\includegraphics[width=\linewidth]{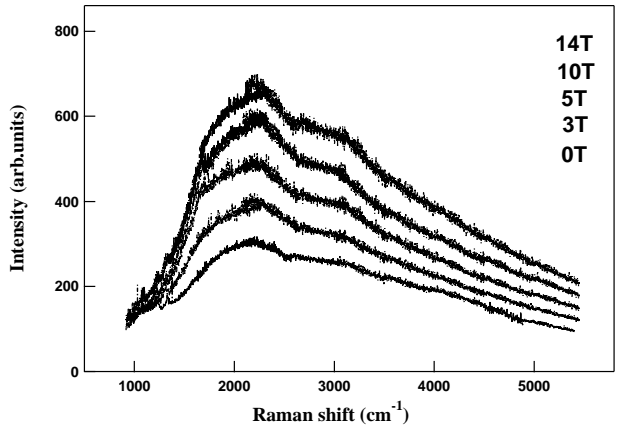}}
\caption{Magnetic field dependence of the ${\rm B_{1g}}$ Raman
response in ${\rm La_{1.88}Sr_{0.12}CuO_{4}}$ at 5 K. The entire
spectrum shown is dominated by the two-magnon response.}
\label{fig1}
\end{center}
\end{figure}

\begin{figure}[t]
%h=here, t=top, b=bottom, p=separate figure page
\begin{center}\leavevmode
\center{\includegraphics[width=\linewidth]{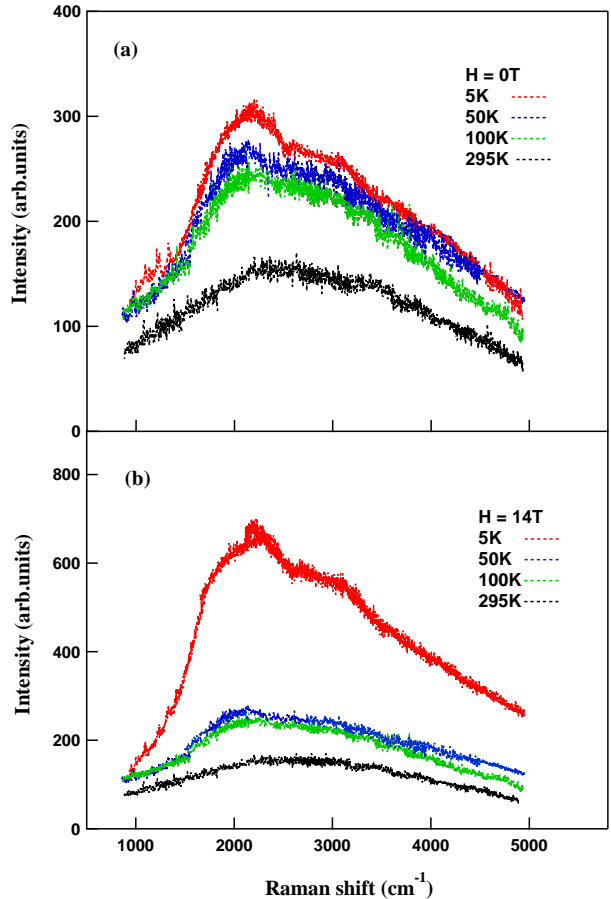}}
%\center{\includegraphics[width=\linewidth]{Figure2a.eps}}
%\center{\includegraphics[width=\linewidth]{Figure2b.eps}}
\caption{(a) Temperature dependence of the ${\rm B_{1g}}$
two-magnon Raman peak in ${\rm La_{1.88}Sr_{0.12}CuO_{4}}$ at (a)
zero field and (b) $\rm H=14$ T.} \label{fig2}
\end{center}
\end{figure}

The central result of this paper is the large magnetic
field-induced intensity enhancement of the low temperature
two-magnon Raman spectrum shown in Fig. 1. While the two-peak line
shape is only weakly affected by the field, a field of 14T leads
to an intensity increase of more than a factor of two compared to
the zero-field data. Within the experimental error, the field
dependence of the energy-integrated spectral weight is linear
(Fig. 3), in contrast to the sublinear field dependence of the
magnetic Bragg reflections observed in some of the neutron
diffraction experiments \cite{lake_nature,khaykovitch}.

The profound magnetic field induced renormalization of the
high-energy magnon spectrum is unexpected. Several cross checks
were performed in order to definitively rule out experimental
artefacts. First, the experiment was repeated at T = 50K, 100K,
and room temperature, where the spectra were found to be field
independent within the experimental error. This is consistent with
the temperature dependence of the static magnetic order determined
in the neutron diffraction experiments
\cite{katano,lake_nature,khaykovitch}. Second, the intensities of
several ${\rm A_{1g}}$ phonon modes were monitored as a function
of magnetic field, and no field dependence was observed within the
experimental error.

Third, the experiment was repeated under the same conditions on an
insulating, antiferromagnetically ordered ${\rm La_{2}CuO_{4}}$
crystal. Since the magnon bandwidth is much smaller than the
Mott-Hubbard gap of 2 eV, the field affects the electrons only via
the Zeeman term $g \mu_B H$ in the Hamiltonian. In a 14T field,
this term leads to a splitting of the two degenerate magnon
branches of order 10 cm$^{-1}$, more than two orders of magnitude
lower than the energy of the two-magnon peak. Given the large
intrinsic width of this peak, a field-induced intensity or
lineshape renormalization should thus not be observable for
undoped ${\rm La_{2}CuO_{4}}$. The ${\rm B_{1g}}$ Raman spectra
displayed in Fig. 4 demonstrate that this expectation is indeed
confirmed by our experiment. Within the experimental error, the
two-magnon profile of ${\rm La_{2}CuO_{4}}$ (which again agrees
very well with prior work on this system \cite{sugai}) is
unaffected by a 14T field. These experimental cross checks provide
reassurance that the magnetic field effect we have observed for
the two-magnon peak in ${\rm La_{1.88}Sr_{0.12}CuO_{4}}$ is
genuine.

\begin{figure}[t]
%h=here, t=top, b=bottom, p=separate figure page
\begin{center}\leavevmode
\center{\includegraphics[width=\linewidth]{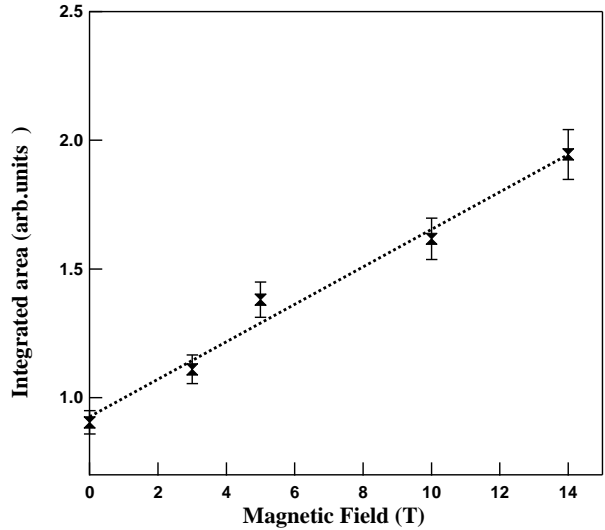}}
\caption{Energy-integrated spectral weight of the two-magnon Raman
peak in ${\rm La_{1.88}Sr_{0.12}CuO_{4}}$ at 5 K, evaluated by
extracting the total areas under the curves of Fig. 1.}
\label{fig3}
\end{center}
\end{figure}

The magnetic field dependence of electronic excitations in high
temperature superconductors has been the subject of several prior
Raman scattering experiments. For instance, recent work on lightly
doped, nonsuperconducting ${\rm La_{1-x}Sr_{x}CuO_{4}}$ with $\rm
x \leq 0.03$ has uncovered a field-induced renormalization of
low-energy magnons with energies of order 10 cm$^{-1}$ (Ref.
\cite{blumberg1}). Since this energy scale is comparable to the
Zeeman energy, these observations are amenable to an
interpretation in the framework of the conventional spin wave
theory, at least on a qualitative level \cite{blumberg1}. An
earlier Raman scattering experiment investigated electronic
excitations in highly overdoped, superconducting $\rm Tl_2 Ba_2
CuO_{6+\delta}$ in fields exceeding the upper critical field
\cite{blumberg3}. As a consequence of the suppression of
superconductivity, the Raman intensity below the superconducting
energy gap, $2 \Delta$, was observed to increase with field, while
that of the broad density-of-states peak above $2 \Delta$ was
reduced, in qualitative agreement with the standard BCS theory of
superconductivity.

The field-induced enhancement of the high-energy Raman intensity
in ${\rm La_{1.88}Sr_{0.12}CuO_{4}}$ defies a description in terms
of such conventional models. As already pointed out above, the
two-magnon peak energy is more than two orders of magnitude larger
than the Zeeman energy in a 14T field. The mechanism invoked to
explain the field-induced renormalization of low-energy magnons in
lightly doped ${\rm La_{1-x}Sr_{x}CuO_{4}}$ (Ref.
\onlinecite{blumberg1}) is therefore not applicable to our system.
Likewise, our results cannot be explained by the BCS theory,
because the energy of the two-magnon Raman peak is much larger
than that of the superconductivity-induced $2\Delta$ peak
\cite{chen,sugai2}. Further, we have shown that its amplitude {\it
increases} with field, in contrast to the {\it decrease} predicted
by the BCS theory and observed in $\rm Tl_2 Ba_2 CuO_{6+\delta}$
\cite{blumberg3}.

\begin{figure}[t]
%h=here, t=top, b=bottom, p=separate figure page
\begin{center}\leavevmode
\center{\includegraphics[width=\linewidth]{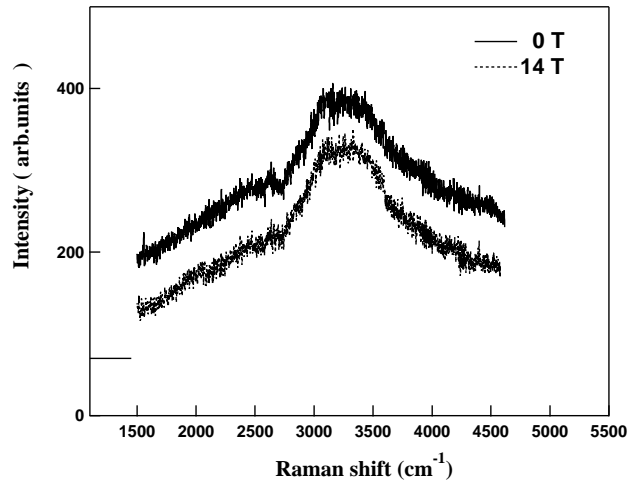}}
\caption{Magnetic field dependence of the ${\rm B_{1g}}$
two-magnon peak in undoped ${\rm La_{2}CuO_{4}}$ at 5 K.}
\label{fig4}
\end{center}
\end{figure}

Since the two-magnon peak is a signature of local
antiferromagnetic interactions, our data are qualitatively
consistent with the notion of a magnetic field-induced enhancement
of antiferromagnetic spin correlations developed on the basis of
low-energy spectroscopy experiments
\cite{katano,lake_nature,khaykovitch,mitrovic,kumagai,kiefl}.
However, they extend the energy range probed by the earlier
experiments by more than an order of magnitude and demonstrate
that a field of 14T profoundly affects the magnon spectrum over
its {\it entire} band width up to the highest-energy, local
spin-flip excitations.

This surprising observation is difficult to reconcile with
theories based on soft-mode behavior controlled by a nearby
quantum critical point \cite{sachdev}. The fact that the
characteristic two-peak lineshape of the two-magnon peak is only
weakly affected by the field, while its amplitude increases
linearly, rather suggests a much simpler picture based on the
coexistence of two phases with very different electronic
structures: a phase with localized electrons and well-developed
local antiferromagnetic order that gives rise to the two-magnon
peak (such as a ``striped" phase); and a phase that gives a much
weaker contribution to the high-energy Raman spectrum (such as a
phase dominated by fermionic quasiparticles that also sustains
superconductivity). The intensity enhancement of the two-magnon
peak then simply reflects a magnetic field-induced increase of the
volume fraction of the former phase. As the temperature is
increased above the critical point for spontaneous phase
separation, the system enters a homogeneous phase without static
magnetic order. The magnetic field effect is hence expected to
disappear, as experimentally observed (Fig. 2). This is not
inconsistent with scenarios in which antiferromagnetic order is
nucleated by vortices
\cite{arovas,ogata,ting,zhang,franz,berlinsky,bishop,hedegard}.
Note, however, that vortices do not seem to be {\it required} for
antiferromagnetic order, as manifestations of static magnetic
order are present even in zero field \cite{katano,suzuki,ohsugi}.
This suggests that the primary effect of the magnetic field is to
shift the thermodynamic balance of the antiferromagnetic and
superconducting phases \cite{sachdev}, not to create the vortices.
Our data are also consistent with a coexistence of superconducting
and magnetically ordered phases on a mesoscopic scale, as observed
for instance for metallic and charge-ordered insulating phases in
some manganites \cite{dagotto}. While our Raman experiments are
sensitive predominantly to regions with magnetic order, recent
infrared experiments on the Josephson plasma resonance in ${\rm
La_{1.875}Sr_{0.125}CuO_{4}}$ have provided complementary evidence
of a spatially inhomogeneous {\it superconducting} state
\cite{basov}. A very recent high-field magnetoresistance study
\cite{ando} has come to a similar conclusion.

In conclusion, we have reported the discovery of a large magnetic
field-induced spectral weight enhancement of the two-magnon Raman
peak in superconducting ${\rm La_{1.88}Sr_{0.12}CuO_{4}}$. These
data are most naturally explained in a two-phase coexistence
scenario, where the magnetic field enhances the volume fraction of
a phase with local magnetic order at the expense of the
superconducting phase. This indicates that these two phases are
separated by a first-order transition, and not by a quantum
critical point.

We acknowledge fruitful discussions with G. Aeppli, C. Bernhard,
and E. Demler. One of us (L.H.M.) acknowledges financial support
from the Alexander-von-Humboldt Foundation.

\end{document}